# Ground-state properties of some N=Z medium mass heavy nuclei

Serkan Akkoyun[1], Tuncay Bayram[2], Şevki Şentürk[3]

[1]*Department of Physics, Faculty of Science, Cumhuriyet University, Sivas, Turkey*
[2]*Department of Nuclear Energy Engineering, Sinop University, Sinop, Turkey*
[3]*Department of Quantum Systems Modelling, Sinop University, Sinop, Turkey*

**Abstract.** The ground-state properties of $^{64}$Ge, $^{68}$Se, $^{72}$Kr and $^{76}$Sr (N=Z) nuclei have been investigated by using Hartree-Fock-Bogolibov (HFB)method with Sly4 Skyrme forces and Relativistic Mean Field (RMF) model with NL3 and recently proposed DEFNE interaction parameters sets. For determination of ground-state axially deformed shape and quadrupole moment constrained calculations have been employed in RMF model. The results of the present study have been compared with each other and available experimental data in the literature. The ground-state binding energies, neutron, proton and charge radii, quadrupole moment deformation parameters of these nuclei have been calculated. Furthermore, neutron skin thickness of considered nuclei as a function of deformation parameter have been obtained and discussed in detail.

**Keywords:** Nuclear properties, neutron skin thickness, HFB method, RMF model, N=Z nuclei

**Introduction**

Atomic nuclei at or around the N=Z line have been of great interest in nuclear physics. In this type of nuclei protons and neutrons occupy same orbits. They form a unique laboratory for studying various phenomena such as the interplay of T=0 and 1 states close to the ground state, the shape coexistence along the N=Z line and the role of neutron–proton pairing correlations. In addition, these nuclei lie along the explosive rp-process nuclear synthesis pathway [1].From theoretical side one of the main aims of research in nuclear physics is to describe the ground-state properties of all nuclei in the periodic table with one nuclear model. Due to the fact that the strong interaction is not yet fully understood and to the numerical difficulties in handling the nuclear many-body problem, relativistic and nonrelativistic phenomenological descriptions have received much attention for describing the ground-state properties of nuclei. Because of this reason Hartree-Fock-Bogoliubov (HFB) method [2,3] and Relativistic Mean Field (RMF) model [4-8] have been used successfully for describing ground-state properties of nuclei during the last decades. Therefore main interest of the study is investigation of some ground-state nuclear properties of $^{64}$Ge, $^{68}$Se, $^{72}$Kr and $^{76}$Sr nuclei.

In the present study we have employed HFB method and RMF model for these nuclei by considering axially deformed case. Because these nuclei do not have shell closure which means that these nuclei possibly have deformed shape in their ground-states. We have calculated binding energies, radii and quadrupole deformation parameters of considered nuclei by using both HFB method and RMF model. The results have been found as in agreement with the available data. Furthermore, in the case of RMF model applications for these nuclei, we have used quadrupole constrained RMF calculations. This application provides ground-state properties of nuclei as a function of deformation parameter ($\beta_2$). By using this method we have carried out neutron skin thickness of $^{64}$Ge, $^{68}$Se, $^{72}$Kr and $^{76}$Sr nuclei as a function of fixed $\beta_2$. We have discussed effect of quadrupole deformation parameter on neutron skin thickness in detail. Also obtained potential energy curves (PECs) have been used for determination of the ground-state shape of these nuclei.

**Material and Methods**

Giving the fully descriptions of HFB and RMF formalism is not suitable in this paper because the complexity of phenomenology of both models. Therefore the authors suggest Ref. [2-8] to read for more information about these models. In the HFB method theory, many properties of nuclei can be described in terms of a model of independent particles moving in an average potential whose space dependence closely follows the matter distribution. In our axially deformed HFB calculations we have fallowed the recipe of Stoitsov et al. [9] and used effective Sly4 Skyrme forces. The number of shells taken into account is 20. We have calculated the ground-state properties of $^{64}$Ge, $^{68}$Se, $^{72}$Kr and $^{76}$Sr nuclei in HFB method both for oblate and prolate shape. The shape of nuclei which gives lowest binding energy for nuclei has been taken as the ground-state of nuclei.

Nucleons interact with each other via exchange of various mesons in the RMF model. It starts with a Lagrangian density includes terms related with free nucleons, free mesons and nucleon-meson interactions. This Lagrangian density includes some adjustable parameters such as masses of nucleons, mases of mesons, coupling constants of nucleon-meson interactions and coupling constants of self-interactions of mesons. These parameters called as RMF parameters and there can be found three types of RMF model parameters in literature. In this study, we have used non-linear version RMF model with NL3 [10] and recently proposed DEFNE parameters [11]. For determination of the ground-state properties of $^{64}$Ge, $^{68}$Se, $^{72}$Kr and $^{76}$Sr nuclei in RMF model we have employed quadrupole moment constrained calculations in such a way that for fixed quadrupole moment we have calculated binding energies of nuclei. Thus we have obtained PECs of considered nuclei. One can easily determine the ground-state of nuclei by investigating of lowest point of PECs. 12 and 20 shells have been considered for fermionic and bosonic expansion in the present study. For numerical explanation of RMF model can be found in Ref. [12].

**Results and Discussions**

The calculated ground state binding energies of $^{64}$Ge, $^{68}$Se, $^{72}$Kr and $^{76}$Sr nuclei by using HFB method with Sly4 method and RMF model with NL3 and DEFNE parameter set generated recently by our working group are listed in Table 1. Also the predictions of finite-range-dropled-model (FRDM) and the experimental data are listed for comparison. The binding energy values of considered nuclei calculated with HFB method and RMF model are close to each other and agreement with experimental data. According to the calculated ground-state binding energies, DEFNE parameter set results are generally closer to the experimental values than the thoseNL3 parameters for the several N=Z nuclei except for $^{64}$Ge. However closest predictions for binding energies are those of FRDM when they compared with those of HFB method and RMF model. It should be noted that the success of FRDM is aroused from the fitting of many parameters by using experimental data. Relatively small numbers of parameters are refitted from experimental data.

**Table 1** The ground state binding energy (in units of MeV) of some N=Z nuclei calculated by the HFB method with the Skyrme force Sly4 and RMF model with DEFNE and NL3 parameters. The predictions of FRDM [13] and experimental data [14] are also shown for comparison.

| Isotope | HFB | RMF-DEFNE | RMF-NL3 | FRDM | Experiment |
|---------|---------|-----------|---------|--------|------------|
| $^{64}$Ge | 543.899 | 528.841 | 538.433 | 544.61 | 545.882 |
| $^{68}$Se | 575.410 | 573.974 | 583.852 | 575.44 | 576.463 |
| $^{72}$Kr | 604.381 | 614.592 | 623.259 | 606.72 | 606.911 |
| $^{76}$Sr | 635.171 | 652.140 | 657.335 | 638.26 | 637.936 |

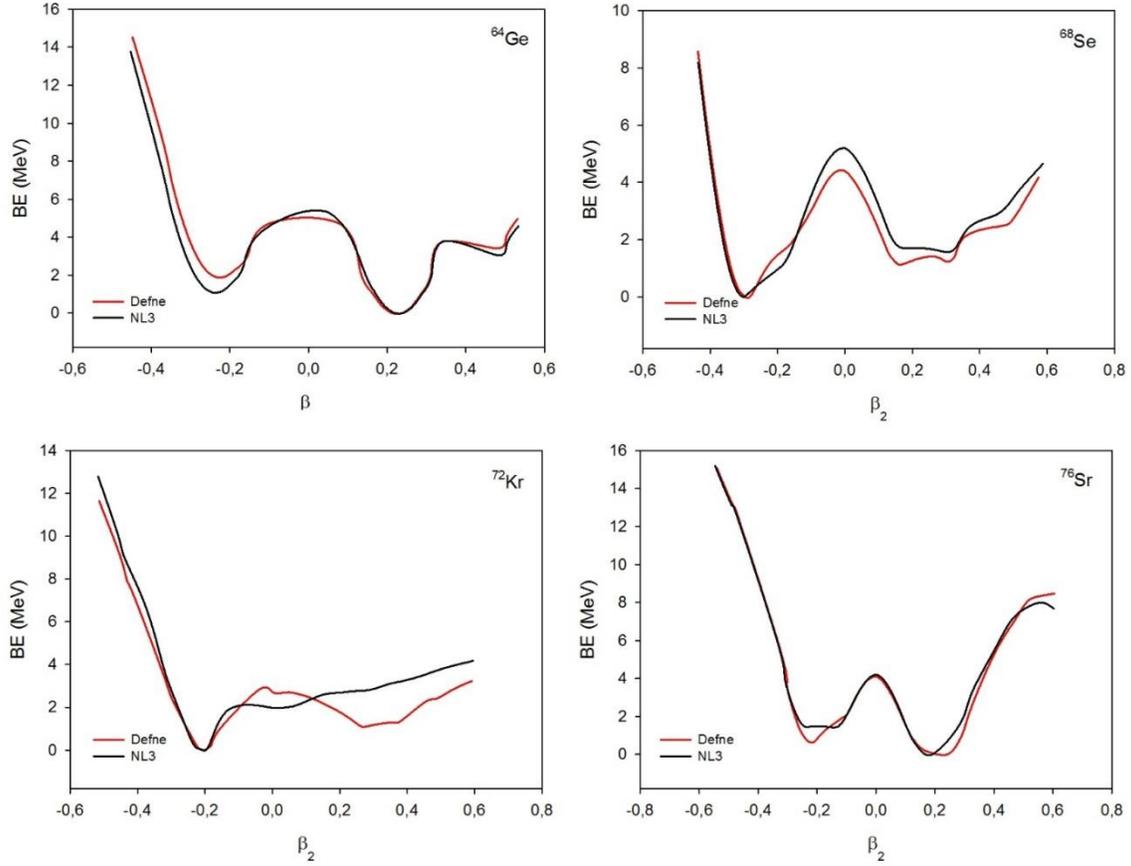

**Fig. 1** The calculated PECs of $^{64}$Ge, $^{68}$Se, $^{72}$Kr and $^{76}$Sr nuclei in the RMF model.

To determine the ground-state shape of nuclei, the PEC is an important tool. In Fig. 1, the calculated PECs of $^{64}$Ge, $^{68}$Se, $^{72}$Kr and $^{76}$Sr nuclei in RMF model with NL3 and DEFNE interactions are shown as a function of $\beta_2$. In this figure lowest binding energy has been taken as a reference. Positive value of $\beta_2$ is related with prolate shape while negative value of $\beta_2$ corresponds with oblate shape. As can be seen in the PECs of Fig. 1, $^{64}$Ge and $^{76}$Sr nuclei have prolate shape whereas $^{68}$Se and $^{72}$Kr have oblate shape. We have also calculated the $\beta_2$ quadrope deformation parameters of the nuclei. Other than $^{68}$Se, were found the nuclei to be compatible with each other and gives the same ground state shapes. In the case of $^{68}$Se, HFB and RMF calculations suggest that the nuclei is oblate shaped in its ground state whereas FRDM result indicates it as prolate. According to the common literature results [15], the nucleus has oblate shape as indicated in the present work (Table 2).

**Table 2** The calculated quadrupole deformation parameter $\beta_2$ for the ground-state of $^{64}$Ge, $^{68}$Se, $^{72}$Kr and $^{76}$Sr nuclei. The FRDM predictions [13] are shown for comparison.

| Isotope | HFB | RMF-DEFNE | RMF-NL3 | FRDM |
|---|---|---|---|---|
| $^{64}$Ge | 0.183 | 0.237 | 0.239 | 0.207 |
| $^{68}$Se | -0.225 | -0.295 | -0.300 | 0.233 |
| $^{72}$Kr | -0.174 | -0.202 | -0.206 | -0.366 |
| $^{76}$Sr | 0 | 0.242 | 0.190 | 0.402 |

An important quantity of nuclear properties of nuclei is root mean square (rms) charge radii ($r_c$). Therefore we have calculated rms charge radii of considered nuclei in both HFB method and RMF models. The calculated charge radii of the nuclei are listed in the Table 3. Also, neutron and proton radii are listed in the table. It should be noted that there is no experimental data for these nuclei in literature. The calculated results from HFB and RMF models are in agreement with each other.

**Table 3** The calculated rms charge radii of considered N=Z nuclei.

| Isotope | HFB | $r_c$(DEFNE/NL3) | $r_n$ (DEFNE/NL3) | $r_p$(DEFNE/NL3) |
|---|---|---|---|---|
| $^{64}$Ge | 4.00 | 3.98/3.97 | 3.83/3.83 | 3.90/3.89 |
| $^{68}$Se | 4.10 | 4.05/4.03 | 3.98/3.99 | 3.95/3.95 |
| $^{72}$Kr | 4.16 | 4.05/4.03 | 4.06/4.08 | 3.97/3.95 |
| $^{76}$Sr | 4.20 | 4.07/4.05 | 4.16/4.18 | 3.99/3.97 |

Neutron skin thickness of nuclei has an important role in theoretical investigations of stellar studies. It can be obtained by using neutron and proton radii of nuclei in the formula $\Delta r_{np} = r_n - r_p$. The calculated $\Delta r_{np}$ by using DEFNE and NL3 effective interactions in RMF model are shown in Fig. 2. As it is expected that $\Delta r_{np}$ is increasing with adding of nucleons.

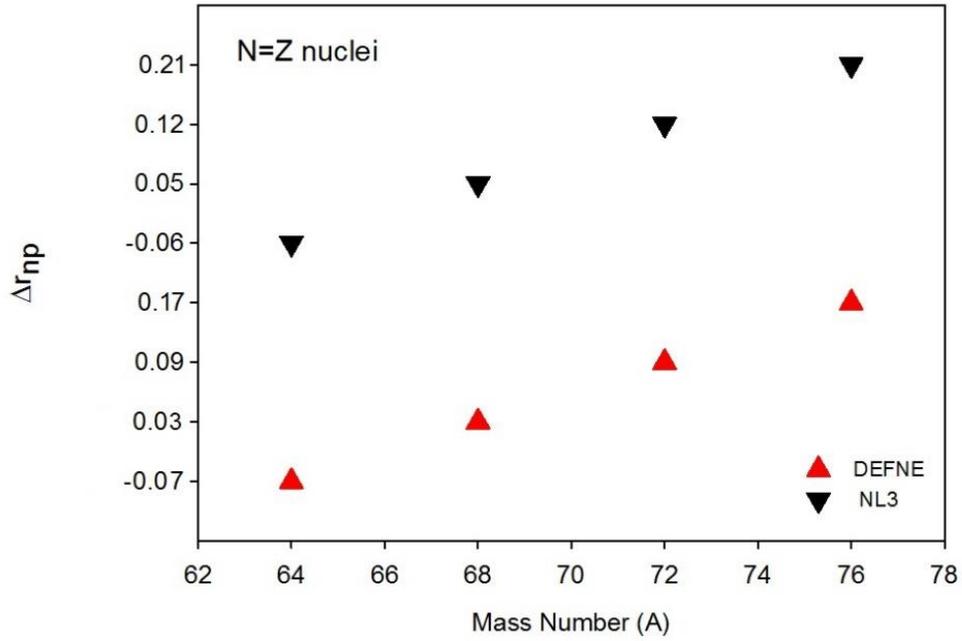

**Fig. 2** Neutron skin thickness of considered N=Z nuclei within the framework of non-linear RMF model.

Furthermore, an investigation of effect of quadrupole deformation parameter on neutron skin thickness can be an interesting study. In Fig. 3, $\Delta r_{np}$ values of $^{64}$Ge, $^{68}$Se, $^{72}$Kr and $^{76}$Sr nuclei calculated by using DEFNE and NL3 parameters sets are shown as a function of $\beta_2$. General tendency of curves are same both for DEFNE and NL3 interactions. However, the NL3 parameter set predictions for neutron skin thickness are somehow higher than those of NL3.

**Conclusions**

We have performed HFB and RMF models for investigation of the ground-state properties of even-even N=Z nuclei with proton numbers Z=32–38. The binding energy, rms charge radius, proton and neutron radius and neutron skin thickness of the nuclei are reproduced well by using the methods. In particular, the DEFNE interaction in the frame of RMF model is found to be a successful interaction for describing of ground-state properties of the nuclei. The ground-state shapes of $^{64}$Ge, $^{68}$Se, $^{72}$Kr and $^{76}$Sr nuclei have been determined by using the PECs of the nuclei by applying quadrupole moment constrained RMF calculations.

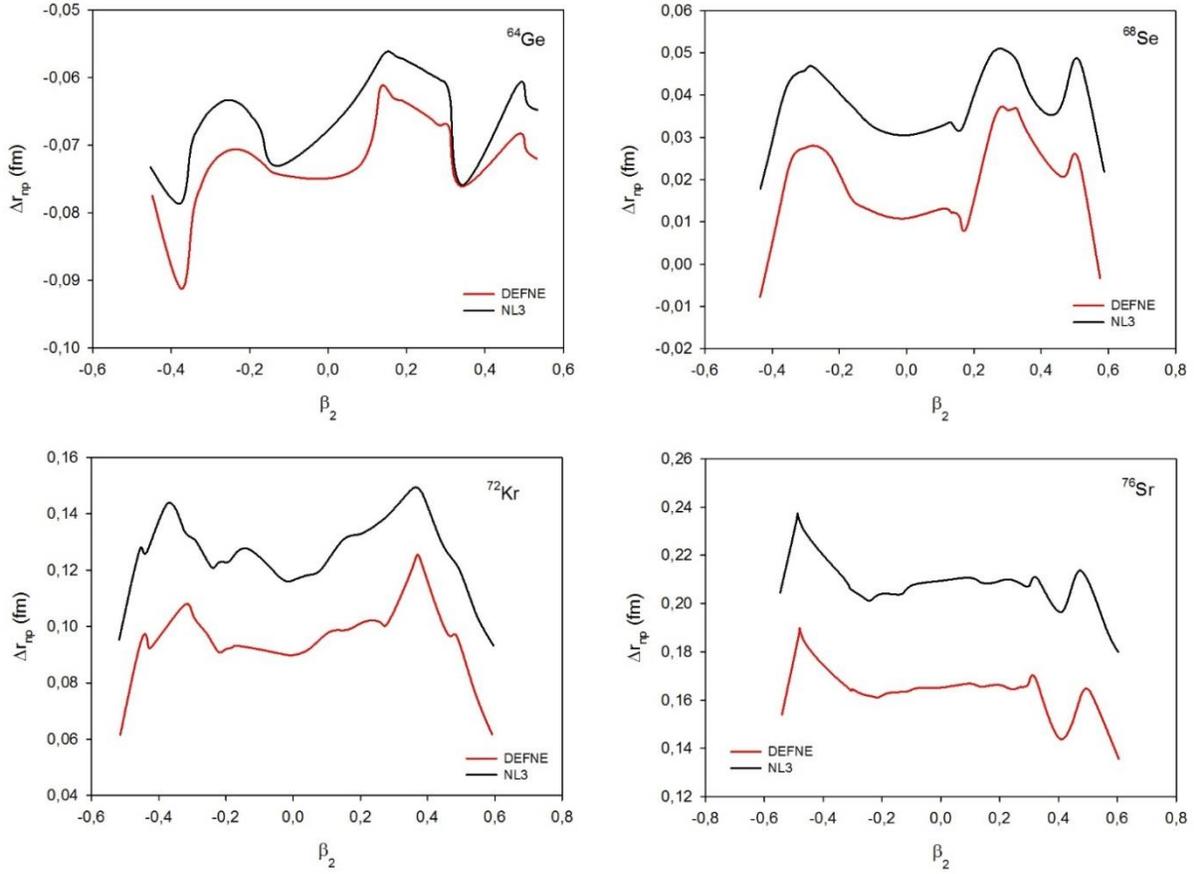

**Fig. 3** The calculated neutron skin thickness of considered N=Z nuclei in RMF model with DEFNE and NL3 interactions as a function of deformation parameter.

**Acknowledgements**

This work was supported by Scientific and Technological Research Council of Turkey (TÜBİTAK) with Project no. 115F291 and Cumhuriyet University Scientific Research Center with project number SHMYO-10.